# Rad-hard vertical JFET switch for the HV-MUX system of the ATLAS upgrade Inner Tracker

Pablo Fernández-Martínez,[a,*] Miguel Ullán,[a] David Flores,[a] Salvador Hidalgo,[a] David Quirion,[a] and David Lynn[b]

[a] *Instituto de Microelectrónica de Barcelona, IMB-CNM (CSIC),*
   *Campus UAB, E-08193, Bellaterra (Barcelona), Spain*

[b] *Brookhaven National Laboratory,*
   *Upton (NY), USA*
   *E-mail*: pablo.fernandez@csic.es

ABSTRACT: This work presents a new silicon vertical JFET (V-JFET) device, based on the trenched 3D-detector technology developed at IMB-CNM, to be used as switches for the High-Voltage powering scheme of the ATLAS upgrade Inner Tracker. The optimization of the device characteristics is performed by 2D and 3D TCAD simulations. Special attention has been paid to the on-resistance and the switch-off and breakdown voltages to meet the specific requirements of the system. In addition, a set of parameter values has been extracted from the simulated curves to implement a SPICE model of the proposed V-JFET transistor. As these devices are expected to operate under very high radiation conditions during the whole experiment life-time, a study of the radiation damage effects and the expected degradation on the device performance is also presented at the end of the paper.

# Contents

## 1. Introduction

The upgrade of the ATLAS tracker for the future High-Luminosity LHC (HL-LHC) will demand a substantial improvement of both the radiation sensors and their associated electronics, in terms of performance, radiation hardness, and compactness [1]. Specifically, the upgrade of the silicon strip sensors represents a large increase in area and in the number of channels. Cable space limitations do not permit each sensor to be individually biased. Therefore, collections of sensors will need to share a common high voltage bias.

A High-Voltage Multiplexing (HV-MUX) scheme has been proposed for the ATLAS ITk (Inner Tracker) [2] in which a failing sensor can be disconnected from the bias bus in order to permit normal operation of the remaining sensors. To this end, each sensor requires a slow-controlled switch that can survive the high radiation environment and operate in a high magnetic field, as well as being capable of switching more than 500 V for the lifetime of the HL-LHC. Commercial switches, including Gallium Nitride (GaN) transistors, Silicon JFET, Silicon MOSFET and Silicon Carbide (SiC) JFET devices are being evaluated in the strip community [2], although not a fully satisfactory component has been found so far.

A new silicon vertical JFET (V-JFET) switch, based on the trenched technology developed at IMB-CNM for 3D radiation detectors [3, 4], is presented in this work as an alternative to the commercial switches. Indeed, the V-JFET is custom designed in order to fulfill the ITk HV-MUX specifications. Among its main features stands its operation in depletion mode (normally-on transistor), which means that it does not require a voltage on the gate electrode for the sensors to be biased. As it is a 3D device with vertical conduction, high voltage capability and low switching voltage are achieved, together with high radiation hardness in terms of ionization damage. Additionally, the layout configuration and use of a p-type substrate will contribute to reduce the effects of radiation induced displacement damage. Finally, it has to be noted that the V-JFET is conceived as a cellular device with an adaptable number and arrangement of the individual cells, in such a way that both the current capability for a proper sensor bias and the area restrictions in the HV circuit board can be fulfilled.

## 2. Device concept

The V-JFET, integrated on a high-resistivity p-type substrate, is conceived as a cellular device, with an adaptable number of cells, typically in the order of several thousands, as depicted in Figure 1-a, for a 5×5 mm² device. Each cell (see Figure 1-b) presents a conduction channel, surrounded by a deep trench with circular or hexagonal patterns. The trenches, generally less than 100 µm deep are filled with highly doped n-type polysilicon to form the gate electrode. The source electrode is implemented with a highly-doped p-type diffusion localized at the center of each cell. The drain electrode is performed on the backside with a blanket boron implant, far away from the trenches, in order to create a wide drift region which allows a voltage capability higher than 1000 V.

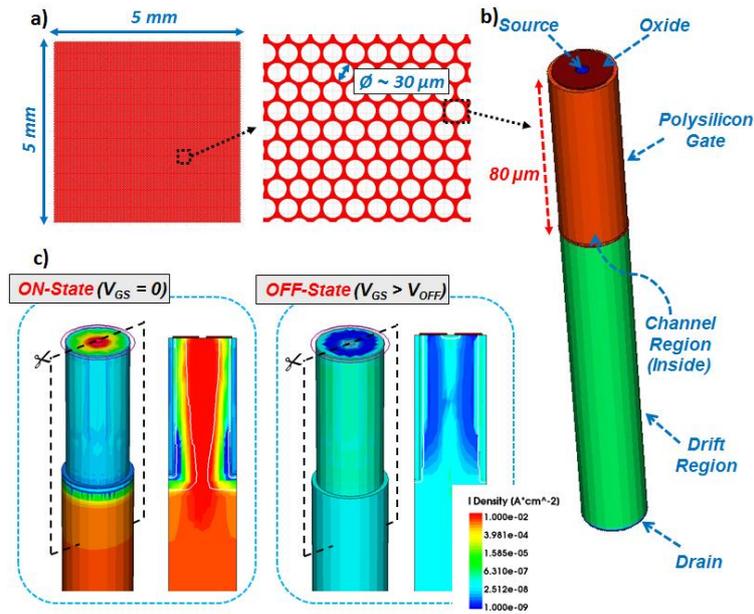

**Figure 1:** V-JFET views: a) Cellular arrangement of the full device with a zoom of circular cells; b) 3D pot of an individual cell; c) Simulated current density through the channel and drift regions in on-state ($V_{GS} = 0$ V) and of-state ($V_{GS} > V_{OFF}$) conditions.

The device operation is presented in Figure 1-c, where the simulated current density is depicted for the two operating modes. In normal operation, Gate-to-Source voltage ($V_{GS}$) is held at 0 V, allowing a current flow through the cell channel. If $V_{GS}$ is increased above a threshold value ($V_{OFF}$), the channel becomes fully depleted and the V-JFET turns-off. A certain level of leakage current is still flowing through the channel in the off-state, but its value is several orders of magnitude lower than in the on-state, thus ensuring a proper switching performance. This behavior can be alternatively observed in Figure 2-a, where the simulated transfer characteristic curve ($I_{DS}$ vs $V_{GS}$) of the V-JFET is represented in logarithmic scale. The simulated device in this example has 10000 parallel cells, with 35 µm channel diameter and a gate trench depth of 80 µm. The doping concentration of the p-type silicon substrate has been set to $2\times10^{13}$ cm$^{-3}$, which approximately corresponds to a resistivity of 650 Ω·cm. With this set of values, the V-JFET is able to drive more than 6 mA in the on-state, whereas the off-state leakage current in the range of a few nA, and the switch-off voltage ($V_{OFF}$) is lower than 2 V. The output characteristic ($I_{DS}$ vs $V_{DS}$, for $V_{GS} = 0$V), is shown in Figure 2-b. As $V_{DS}$ is raised, the channel becomes increasingly depleted at the bottom side whereas the top side remains undepleted.

Current flow is still possible but it slowly increases with $V_{DS}$ until it reaches a saturation value, when the bottom of the channel eventually becomes pinched-off. Correspondingly, three different conduction modes can be identified in the output characteristic, as in conventional JFET transistors [5]: linear, while the whole channel remains undepleted; triode, as the bottom of the channel becomes progressively depleted; and saturation, once pinch-off is reached.

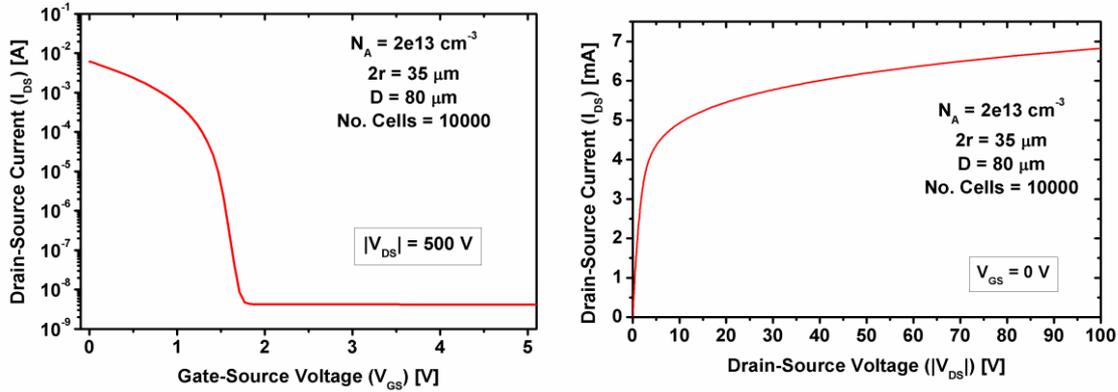

**Figure 2.** 3D-simulated characteristic curves for a V-JFET with $10^4$ cylindrical cells: a) $I_{DS}$ vs. $V_{GS}$ curve for $|V_{DS}| = 500$ V; b) $I_{DS}$ vs. $V_{DS}$ curve for $V_{GS} = 0$ V.

For the specific application in the HV-MUX system, $V_{DS}$ will be determined by the sensor current, in such a way that the demanded current will set the corresponding $V_{DS}$ value, according to the output characteristic represented in Figure 2-b. This value is defined as the voltage drop ($V_{drop}$), as it stands for the reduction in the voltage bias of the sensor. In this sense, $V_{drop}$ is intended to remain as low as possible, and the V-JFET should be preferably operated in the linear mode. If an abnormally high current value is demanded from the sensor, the V-JFET will be lead to the triode or saturation modes, with a significant increase in $V_{drop}$. At that point, the control electronics will be devised to increase $V_{GS}$ above the $V_{OFF}$ value, thus switching-off the transistor. Nevertheless, it is worth noting that the V-JFET might be used as well as a current limiter, since even if $V_{GS}$ were maintained at 0 V, the transistor would not be able to give a current value above the saturation level.

## 3. Device optimization for the ITk HV-MUX requirements

The specific use of the new V-JFET on the ITk HV-MUX system delimits the device performance. In this sense, the optimization of the design is focused on several figures-of-merit, which are specifically defined for this application. Table 1 compiles the specifications for such figures-of-merit, together with the target values aimed at this study. The mentioned figures-of-merit are:

- **ON-state current ($I_{ON}$):** Defined as the maximum current value required during the on-state operation. It corresponds to the expected value for the leakage current of the irradiated sensors, i.e. 1-10 mA. The target is $I_{ON} = 5$ mA.
- **OFF-state current ($I_{OFF}$):** Defined as the maximum tolerable current for the off-state operation. It should not be greater than 200 µA before irradiation. A target of 10 µA has been fixed. After irradiation the target value has been set to 1 mA.
- **Switch-off Voltage ($V_{OFF}$):** Defined as the $V_{GS}$ value required to obtain a leakage current through the channel ($I_{DS}$) lower than $I_{OFF}$ (i.e. 10 µA). Since the switching performance is controlled with low-voltage electronics, a $V_{OFF} < 3$ V is preferred, whereas the optimal value aims at less than 1 V for non-irradiated devices. This low value has been intentionally chosen, as a certain increase is expected after irradiation.

- **Voltage drop ($V_{drop}$):** Defined as the $V_{DS}$ value corresponding to $I_{ON}$ (i.e. 5 mA) in the output characteristic. As long as the device is operating on the linear mode, $V_{drop}$ is completely determined by the on-resistance ($R_{ON}$). However, $V_{drop}$ will grow up very fast if the V-JFET enters the triode or saturation modes. As a general rule, $V_{drop}$ has to be as low as possible, being targeted at < 3 V.
- **Drain-to-Gate breakdown Voltage ($V_{br\text{-}DG}$):** Corresponds to the $V_{DS}$ value for $I_G = 10$ µA, which is a value 1000 times higher than the typical $I_G$ value in non-irradiated devices. This value determines the voltage capability of the device, which is required to be higher than 500 V. In this sense, a target value of 1000 V is fixed.
- **Source-to-Gate breakdown Voltage ($V_{br\text{-}SG}$):** Corresponds to the $V_{GS}$ value for $I_G = 10$ µA. In this case, the $V_{br\text{-}SG}$ value is less limiting for the V-JFET performance than that of $V_{br\text{-}DG}$, as the maximum value required for $V_{GS}$ will always remain below 5 V, far enough from the typical $V_{br\_SG}$ which is aimed at ~100 V.

| Figure of Merit | Requirement | Target |
|---:|---|---|
| $V_{br}$ ($V_{br|G\text{-}D}$) | > 500 V | ~ 1000 V |
| $I_{ON}$ | 1-10 mA | 5 mA |
| $V_{drop}$ | few volts | < 3 V |
| $V_{OFF}$ | < 3 V | < 1 V |
| $I_{OFF}$ | < 200 µA | < 10 µA |
| Area | < 9×9 mm$^2$ | 6×6 mm$^2$ |

**Table 1.** Specifications and target values for the main V-JFET figures-of-merit

The selection of the total area of the V-JFET is a challenge. On the one side, large area is required in order to meet the $I_{ON}$ requirement, without leading the transistor into the triode or saturation modes. On the other side, the space reserved to the switch in the HV-MUX system is limited to a maximum area of 9×9 mm$^2$. Hence, the area of the optimized device has been initially fixed to 6×6 mm$^2$.

The figures-of-merit will have different values, depending on the specific design of the V-JFET. In this sense, several design parameters can be tuned to obtain an optimal configuration. The most relevant are shown in Figure 3, on a schematic cross section of a single V-JFET cell. The optimization is focused on the influence of the trench depth (D) and the channel diameter (2r) on the output characteristics, together with that of the substrate doping concentration ($N_A$). Other parameters, such as the trench width (W) or the substrate thickness (T), have been respectively fixed to 5 µm and 300 µm, for technological reasons. The drift region depth (d) is crucial to define the voltage capability of the device; however, it is fully determined by D, once T has been fixed. Finally, two additional design parameters are related with the full device arrangement: the distance between adjacent cells (S) and the number of parallel cells ($N_{cells}$), which determines the total current capability of the device. Adjacent cells can be arranged without separation, by sharing the same gate trench. Nevertheless, an S of few microns is preferred in order to ensure a good technological yield. No significant effect is expected on the full device performance, other than the increase in the area of the individual cells. In the same way, the number of cells is limited by the total available area for the V-JFET. All simulations have been done considering that the full device contains 10000 cells, which is compatible with a device total area of approximately 6×6 mm$^2$ and S = 10 µm.

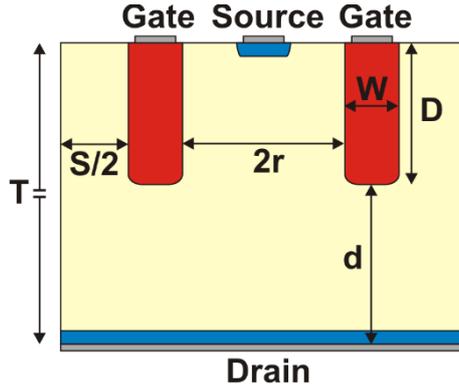

**Figure 3**. Schematic cross section of a single V-JFET cell, with the main design parameters.

The first parameter to be optimized is the trench depth. As shown in Figure 4, the D value affects both $V_{OFF}$ and $V_{br-DG}$. On the one hand, if trenches are made too shallow, $V_{OFF}$ rises up to values above 5 V, which are not suitable for the HV-MUX application. On the other hand, if the trenches are etched very deep in the substrate, the drift region will be shortened and $V_{br-DG}$ will be drastically reduced. Besides, very deep trenches imply technological risks and the consequent reduction in the manufacturing yield. Optimal values for both $V_{OFF}$ and $V_{br-DG}$ can be obtained with D values between 60 and 100 μm. In this sense, an optimal value of D = 80 μm is considered for the following discussions.

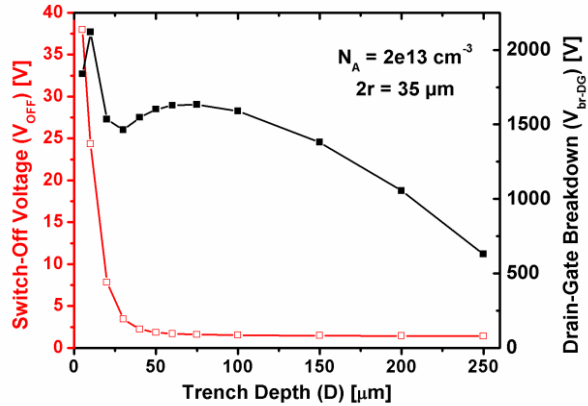

**Figure 4.** $V_{OFF}$ and $V_{br-DG}$ as a function of D, for a V-JFET cell with 2r = 35 μm and $N_A = 2\times10^{13}$ cm$^{-3}$.

Next, $V_{drop}$ and $V_{OFF}$ were evaluated as a function of 2r and $N_A$. Following the results reported in Figure 5, the lower the doping concentration of the channel, the higher the voltage drop: as the channel becomes more resistive, the V-JFET drives less current to the sensor, so $V_{drop}$ is increased. It is worth noting that the $V_{drop}$ increment is not linear with decreasing $N_A$, since, as $N_A$ decreases, the required $V_{DS}$ to drive 5 mA to the sensor is shifted from the linear towards the saturation conduction modes. On the contrary, as $N_A$ is increased $V_{OFF}$ becomes higher, since the depletion of the channel region requires higher voltage on the gate electrode. For the HV-MUX application, both $V_{drop}$ and $V_{OFF}$ are expected to be limited to a few volts (< 3 V, according to Table 1). In this sense, $N_A$ should be preferably limited to values between $1\times10^{13}$ and $5\times10^{13}$ cm$^{-3}$.

At the same time, the channel width (2r) has also an important impact on $V_{drop}$ and $V_{OFF}$ values. As plotted in Figure 5, the wider the channel the higher the voltage required to switch off the transistor. Hence, 2r has to be as low as possible. Nevertheless, for a given $N_A$ value, as the channel becomes narrower, $V_{drop}$ rises. It is even possible that for a 2r value too small, the V-JFET would not be able to meet the $I_{ON}$ = 5 mA target for the ON-state operation, unless a very large area is used. Besides, a technological limitation arises when a channel narrower than 20 μm is intended to be made. Accordingly, 2r values between 20 and 35 μm are preferably chosen for an optimized device.

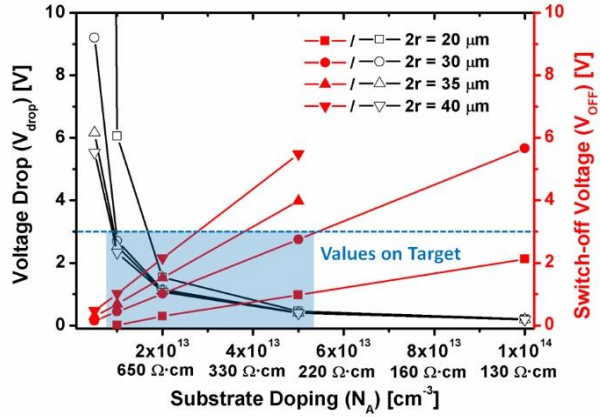

**Figure 5.** $V_{drop}$ and $V_{OFF}$ as a function of $N_A$, for various V-JFET cells with different 2r.

The impact of $N_A$ and 2r on $V_{br-DG}$ and $V_{br-SG}$ has been also evaluated. Whereas changes in 2r have a minor relevance for the final value of these figures-of-merit, their dependence with $N_A$ can be observed in Figure 6. As can be seen, $V_{br-DG}$ experiences a drastic reduction as $N_A$ is increased, since a slower spread of the depletion through the drift region takes place. On the contrary, the reduction in $V_{br-SG}$ with increasing $N_A$ is much smaller. In any case, the targeted range for $N_A$ between $1\times10^{13}$ and $5\times10^{13}$ cm$^{-3}$ (see Figure 5) ensures a safety margin for the optimized V-JFET to meet the specifications.

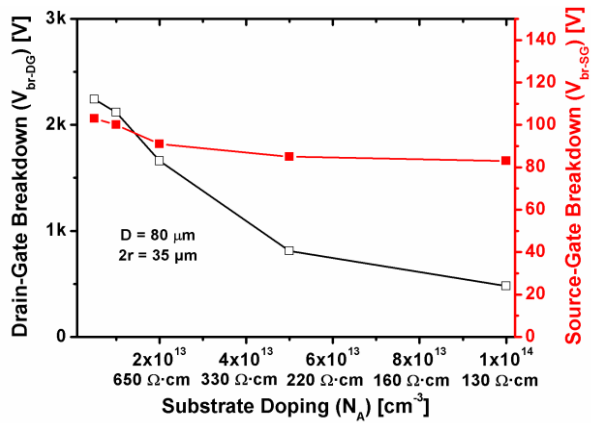

**Figure 6**. $V_{br-DG}$ and $V_{br-SG}$ as a function of $N_A$, for a V-JFET cell with D=80 μm, 2r=35 μm.

Finally, the leakage current through the gate electrode ($I_G$) has been always observed below 10 nA, both under the on-state and the off-state simulated conditions.

## 4. SPICE model

Based on the standard SPICE model for a conventional JFET [6], a set of parameter values has been extracted from the simulated curves to model the performance of the proposed V-JFET. Each cell of the device can be considered as a conventional JFET, corresponding to the channel, connected in series with a high valued resistor, which accounts for the drift region. In this sense, each cell should be modeled considering four parameters, BETA, VTO, LAMBDA and RD. Besides, the off-state performance is adjusted with the Is, Isr, N, and Nr parameters, which model the leakage currents observed in the transistor. Finally, the model is scaled with the AREA parameter to take into account the total number of cells contained in the full device. The extracted values for the relevant parameters are listed in Table 2.

| Parameter | Value | Units |
|---:|---|---|
| BETA | 2.32e-7 | $A \cdot V^{-2}$ |
| VTO | -1.88 | V |
| LAMBDA | 4.31e-3 | $V^{-1}$ |
| RD | 1e6 | Ohm |
| RS | 1e-6 | Ohm |
| Is | 1e-14 | A |
| Isr | 2e-13 | A |
| N | 1 | |
| Nr | 2 | |
| [AREA] | 9500 | |

*Table 2: Values for the SPICE model*

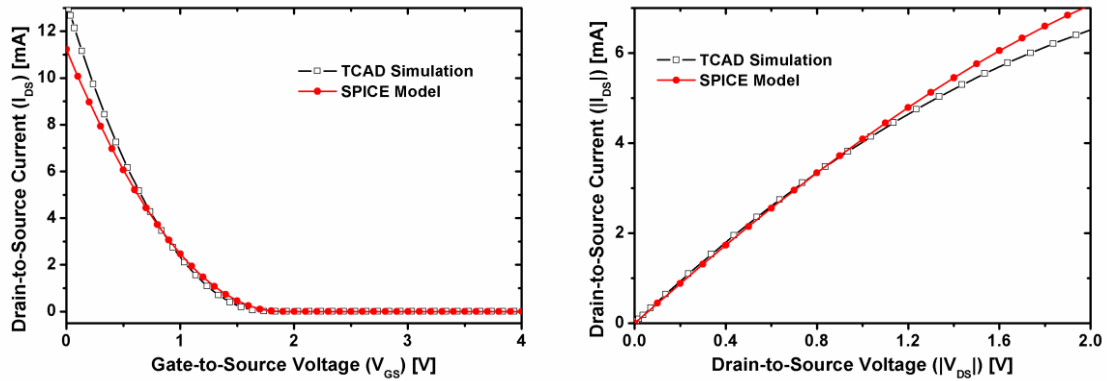

*Figure 7: Transfer characteristic curve (a) and Output characteristic curve (b), in the linear mode, for the SPICE model listed in Table 2, compared with the values simulated with Sentaurus TCAD.*

As shown in Figure 7, the proposed set of values is able to replicate the transfer characteristic (a) and the output characteristic in the linear mode (b). However, it turns out to be not very precise when the triode or the saturation modes are concern. In this sense, a specific SPICE model for the V-JFET transistor, based on measured curves, will be proposed once the first prototypes will be fabricated. At this point, the proposed model is accurate enough for proper simulation of the HV-MUX circuit.

## 5. Radiation Hardness

The application for the HL-LHC requires the V-JFET to be operative under harsh radiation conditions with accumulated hadron fluence up to $2 \times 10^{15}$ $n_{eq}/cm^2$ and total ionizing dose (TID) up to 50 Mrad. The particular design of the V-JFET makes the device hard against TID effects, since the oxide layers, which are the most susceptible regions to accumulate TID induced charges, remain far from the sensitive areas of the device. As a result, minor effects on $V_{OFF}$, $V_{drop}$, $I_{ON}$ or $I_{OFF}$ are expected. This was confirmed by TCAD simulations, where no significant changes in the mentioned figures-of-merit have been observed. Certain degradation on $V_{br-SG}$ is foreseeable. However, the typical operating conditions of the switch, with $V_{GS} < 5$ V, give a safety margin even with the device irradiated up to the highest expected TID. In addition, as long as cells are individually considered, no significant effect on $V_{br-DG}$ is observed in the simulation results. Nevertheless, the accumulation of fixed charge in the oxide at the device periphery might degrade the quality of the edge termination structures included in the full device design. In this sense, mitigation techniques, such as including a p-spray [7], have been implemented for the fabrication.

On the other hand, displacement damage effects are a possible concern for this device. Even manufactured on highly resistive p-type substrate, notable changes can be expected both during the off-state and the on-state operation. In the off-state, the generation of charge carriers, the reduction of minority carrier lifetime and the trap assisted tunneling are major issues, which lead to a rise in the leakage currents ($I_{OFF}$ and $I_G$) together with an increase in $V_{OFF}$. These effects can be studied taking advantage of the simulation techniques developed for evaluating the radiation hardness in silicon detectors [8]. First simulation results show a relevant increase in the leakage current. However, the higher substrate dose used in the V-JFET with respect to that of the strip sensors reduces the relevance of this effect, thus making the V-JFET transistors still usable as switches for the HV-MUX system.

For the on-state operation, the majority carrier removal and the mobility degradation may lead to a reduction of $I_{ON}$, with the corresponding increment in $V_{drop}$, as the operational point is displaced into the triode or saturation conduction modes. A quantitative evaluation of this effect is still under study although previous results on undepleted silicon have shown that the current degradation can be quite limited [9]. In this sense, an irradiation campaign is planned to evaluate the mobility degradation on highly resistive p-type silicon.

## 6. Conclusion

This work presents a new silicon vertical JFET (V-JFET) technology developed at the IMB-CNM (CSIC) for the HV-MUX switches required in the Inner Tracker of the future high luminosity upgrade of the ATLAS experiment. These V-JFET transistors, based on a 3D trenched detector technology, have been optimized to meet the high voltage, low resistance, low switch-off voltage and radiation hardness requirements of the specific application. In this sense, the most relevant figures-of-merit that describe the device performance (i.e. $I_{ON}$, $I_{OFF}$, $V_{OFF}$, $V_{drop}$, $V_{br-DG}$, and $V_{br-SG}$) have been evaluated as a function of several design parameters (i.e. $N_A$, 2r, and D).

In addition, a set of values have been extracted to produce a SPICE model of the V-JFET. Based on the conventional model for JFET transistors, the proposed SPICE model reproduces the transfer and output characteristics in the linear conduction mode, obtained by TCAD simulations. A more complete model will be generated from measured curves on the first prototypes fabricated at the IMB-CNM clean room.

Radiation hardness is now under study with the aid of both TCAD simulations and physical models reported in the literature. The first simulation results confirm the expected hardness against ionization damage, whereas the performance degradation caused by displacement damage is lower than that observed in the associated sensors. In any case, a thorough irradiation program is planned for the fabricated prototypes.

**Acknowledgments**

This work is supported and financed by the Spanish Ministry of Economy and Competitiveness through the Particle Physics National Program (ref.FPA2012-39055-C02-02 and FPA2014-55295-C3-2-R) and co-financed with FEDER funds.